\documentclass[aps,prx,twocolumn,superscriptaddress,showpacs]{revtex4-1}
\usepackage{amsmath,amssymb,graphics,epsfig,epstopdf,color,verbatim,ulem,braket,tabularx}
\usepackage{multirow}
\usepackage[colorlinks,linkcolor=blue,citecolor=blue,urlcolor=blue,bookmarks=false]{hyperref}

\usepackage{listings}
\usepackage{cancel}
\usepackage{mathrsfs}
\usepackage{soul}
\usepackage{color}
\usepackage{url}

\begin{document}

\title{Instability of Pa$\overline 3$ Cs$_{3}$C$_{60}$ at ambient pressure and superconducting state \\ of the FCC phase}

\author{Changming Yue}
\email{changming.yue@unifr.ch}
\affiliation{Department of Physics, University of Fribourg, 1700 Fribourg, Switzerland}

\author{Yusuke Nomura}
\affiliation{Department of Applied Physics and Physico-Informatics, Keio University, 3-14-1 Hiyoshi, Kohoku-ku, Yokohama, 223-8522, Japan}

\author{Kosmas Prassides}
\affiliation{Department of Physics, Graduate School of Science, Osaka Metropolitan University, Osaka 599-8531, Japan}

\author{Philipp Werner}
\email{philipp.werner@unifr.ch}
\affiliation{Department of Physics, University of Fribourg, 1700 Fribourg, Switzerland}

\begin{abstract}
The alkali-doped fulleride Cs$_{3}$C$_{60}$, crystallized in the space group Fm$\overline 3$m or Pm$\overline 3$n, exhibits unconventional $s$-wave superconductivity under pressure with a maximum $T_c\sim 38$ K. 
Recently, 
a new primitive-cubic-structured Cs$_{3}$C$_{60}$ phase 
corresponding to the space group Pa$\overline 3$ has been reported (arXiv:2208.09429) and the authors observed  superconductivity at ambient pressure.  
Using density-functional theory (DFT) calculations, we show that the proposed Pa$\overline 3$ structure 
is not stable under ionic relaxation, but transforms into the FCC structure. 
We study the normal and superconducting state of the stable FCC phase at different temperatures and volumes using 
DFT plus dynamical mean-field theory (DFT+DMFT) in the Nambu formalism. As temperature increases,  the transition between superconductor and normal metal (Mott insulator) at small (big) volume is found to be second (first) order. The recently developed maximum entropy analytic continuation method for the anomalous-self-energy is used to study the momentum-resolved spectra and optical conductivity. 
\end{abstract}
\maketitle

\newpage

{\it Introduction.} \ 
The trivalent alkali-doped fullerides A$_3$C$_{60}$ ($A$ = K, Rb) are 
fullerene-based molecular superconductors at ambient pressure, while the 
volume expanded Cs$_3$C$_{60}$ with larger cation size becomes superconducting
upon application of pressure
 \cite{Gunnarsson1997,Capone2009}.
Cs$_3$C$_{60}$ is known to exist in two types of crystal structures, the FCC (A15) structure
with the space group Fm$\overline 3$m (Pm$\overline 3$n), where the C$_{60}$ molecules
are arranged on a face (body) centered cubic [FCC (BCC)] lattice.
Both FCC and A15 Cs$_3$C$_{60}$ are Mott insulators at ambient pressure. 
The highest $T_c$ in A$_3$C$_{60}$ is realized in A15 Cs$_3$C$_{60}$ at a pressure of $\sim$ 0.7 GPa 
($T_c\sim$ 38 K), while the highest $T_c$ in FCC Cs$_3$C$_{60}$ is 35 K at the same pressure \cite{PalstraCs3C60Tc40K,Ganin2008,Takabayashi2009}.
For both structures, 
the phase diagram in the space of temperature and pressure (or inverse volume per C$_{60}$) exhibits an $s$-wave superconducting state with  
a dome-shaped $T_c$ 
next to an antiferromagnetic or paramagnetic Mott insulator (N\'eel temperature $T_N\sim 46$~K for A15 and 2.2~K for FCC)
\cite{Ganin2008,Takabayashi2009,Zadik2015}, which suggests an unconventional pairing mechanism in A$_3$C$_{60}$ \cite{Capone2009,Ganin2008,Crespi2002,Capone2002}. A$_3$C$_{60}$ is a strongly correlated three-orbital electron system, where the bandwidth $W$ of the 
$t_\text{1u}$ bands ranges from 0.3 to 0.5 eV 
and the on-molecule intra-orbital interaction $U$ ranges from 0.8 to 1.1 eV, depending on the 
type of alkali metal and the volume per C$_{60}$ \cite{Nomura2012}. 
The electron-phonon coupling between the extended $t_\text{1u}$ molecular orbitals and the intramolecular Jahn-Teller $H_g$ phonon modes
results in unusual multiorbital interactions, with an effectively negative (antiferromagnetic) Hund coupling $J_{\mathrm{eff}}$  
\cite{Capone2002,Nomura2015}. As a result, the system exhibits a reversed Hund's rule, which favors low spin and angular momentum states. The negative $J_{\mathrm{eff}}$ and strong on-molecule Coulomb interactions are at the origin of the various 
unconventional normal phases realized in A$_3$C$_{60}$, including the Jahn-Teller metal \cite{Zadik2015,Hoshino2017} and 
the $s$-wave superconducting state
\cite{Capone2009,Hoshino2017}. The pairing glue in this 
$J_{\mathrm{eff}}<0$ system is provided  
by enhanced local orbital fluctuations  
\cite{Hoshino2017}.
In the superconducting phase, these fluctuations are proportional to the pairing strength and
peak in the strongly correlated metal regime \cite{Yue2020}.
Superconductivity in fullerides is thus conceptually linked 
to the spin-freezing induced unconventional superconductivity in positive-$J$ systems like uranium and iron-based superconductors \cite{Hoshino2015} and even cuprates \cite{Werner2016}. 

\begin{figure*}[htp]
\includegraphics[clip,width=0.8\paperwidth,angle=0]{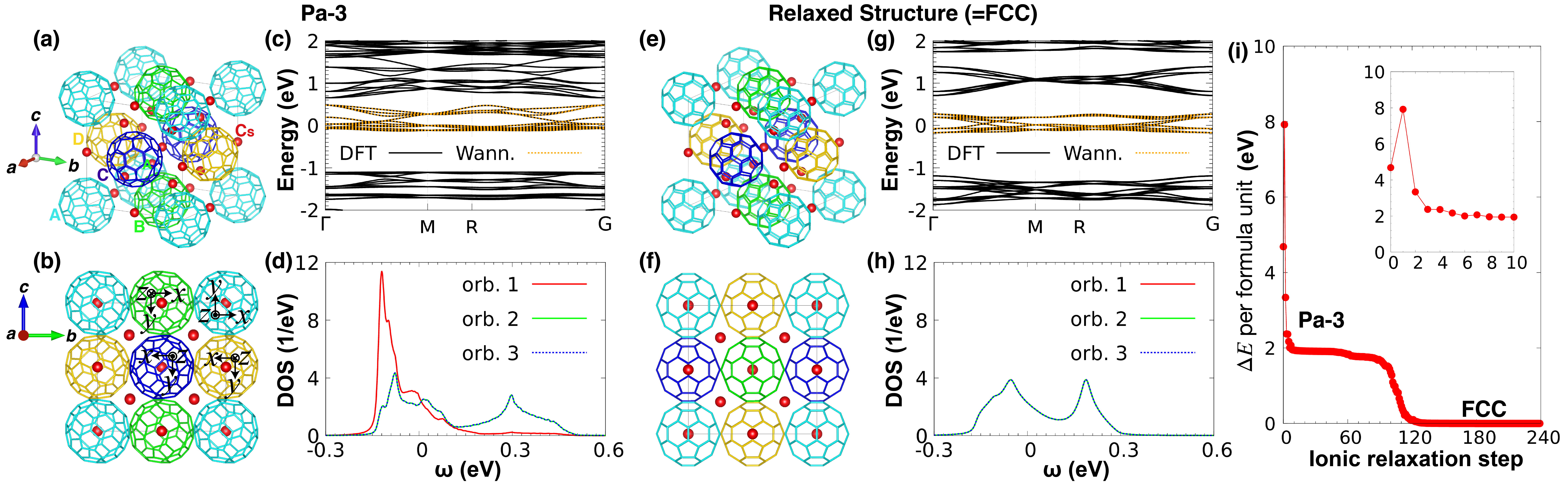}
\caption{Crystal structure and electronic structure of Cs$_3$C$_{60}$. 
(a-b) [(e-f)] Oblique and side view of the Pa$\overline{3}$ [relaxed (FCC)] crystal structure.
The red spheres show the Cs atoms with an exaggeratedly small radius. 
The Pa$\overline{3}$ structure, whose primitive cell is the same as the unit cell, contains four 
C$_{60}$ molecules, with different local coordinates each. 
For a better visualization, molecule A centered at (0.0,0.0,0.0) is colored in cyan, 
B at (0.5,0.5,0.0) in green, C at (0.0,0.5,0.5) in blue, and D at (0.5,0.0,0.5) in gold.
The black arrows in panel (b) illustrate the local $xyz$-coordinate system for each molecule.
The DFT (black) bands and their Wannier interpolations (orange) are shown in panel (c) for the  Pa$\overline{3}$ 
structure and in panel (g) for the relaxed structure. 
Panels (d) and (h) show the corresponding relaxed structure on-molecule orbital-resolved DOS per spin, which are the same 
for all molecules, since they have the same environment relative to the local $xyz$-coordinate system. 
(i) The energy difference $\Delta E$ per C$_{60}$ between the structure under relaxation and
the final stable FCC structure as a function of ionic relaxation steps. 
The initial structure is the experimental Pa$\overline{3}$ structure, whose energy per formula unit is 4.67 eV higher 
than in the final stable FCC structure. The inset shows a zoom of the initial 11 steps.
}
\label{fig:struct}
\end{figure*}

For experiments and practical applications, it is desirable to realize superconductivity at ambient pressure. 
A recent preprint reported superconductivity with $T_c\sim 22$~K in a Cs$_3$C$_{60}$ sample 
at ambient pressure \cite{XJChen2208}. According to this study, the newly discovered phase has a primitive cubic structure 
with the space group Pa$\overline 3$, 
and features a particular type of orientational order of the molecules. 
 Remarkably, the reported evolution of $T_c$ in Pa$\overline 3$ Cs$_3$C$_{60}$ under pressure 
as a function of volume per C$_{60}$ extends to very small volumes ($V=660\sim 700$ \AA$^3$), which has not been reported
before in the FCC and A15 Cs$_3$C$_{60}$ systems \cite{Ganin2008}.  
The trend for very small volumes follows closely that of K$_3$C$_{60}$ under pressure \cite{Zhou1992}. 

Motivated by the findings in Ref.~\onlinecite{XJChen2208}, we investigate the electronic structure and lattice properties of Pa$\overline{3}$ Cs$_3$C$_{60}$. Our simulations based on density functional theory (DFT) show that Pa$\overline{3}$ 
Cs$_3$Cs$_{60}$ at the reported unit cell volume is unstable and relaxes to the FCC structure. 
We thus study the normal and superconducting states of the stable FCC structure for different volumes, but do not find superconductivity for $V \lesssim 700$ \AA$^3$. At higher volumes our simulations yield a 
superconducting phase with $T_c$ values close to those reported in previous experiments on Cs$_3$C$_{60}$ under pressure. 
Also the computed optical conductivity data are consistent with previous optical experiments.

{\it Unstable Pa$\overline{3}$ crystal structure.} \ 
The Pa$\overline{3}$ Cs$_3$C$_{60}$ crystal structure proposed in Ref.~\onlinecite{XJChen2208} is shown in Fig.~\ref{fig:struct}(a,b). In the unit-cell, there are four C$_{60}$ molecules which are orientationally ordered in essentially the same way as in the pure C$_{60}$ crystal \cite{Heiney1991_PRL}. The Pa$\overline{3}$ structure can be constructed from the orientationally ordered FCC one \cite{footnote_disorder}
by rotating the four C$_{60}$ molecules clockwise by the same angle $\phi$ ($\approx 22^\circ$), but about different local $\langle 111\rangle$ axes \cite{Heiney1991_PRL}. 
The Cs atoms (red spheres) are located at the octahedral and tetrahedral holes formed by the close-packed stacking of the C$_{60}$ molecules along the [111] crystal direction. In the local coordinates $\{x,y,z\}$, the four molecules are equivalent to each other, as shown in Fig.~\ref{fig:struct}(b). 

Our analysis suggests that the Pa$\overline{3}$ structure reported in Ref.~\onlinecite{XJChen2208} is not stable. The Cs atoms are easily ionized and donate an electron each to the fullerene molecules. The ionic radius of Cs$^+$ is $r_{\mathrm{Cs^+}}=1.67$~\AA \, and the van der Waals radius of C is $r_{\mathrm{vdW,C}}=1.7$ \AA. However, according to the experimental structure, the nearest Cs$^+$-C distance for the Cs$^+$ located in the tetrahedral hole  is 2.7 \AA, which is much smaller than 
$\sim r_{\mathrm{Cs^+}}$+$r_{\mathrm{vdW,C}}=3.37$~\AA. Such a small Cs-C distance would require a significant external pressure and hence the proposed structure cannot be stable at ambient pressure. 
Pa$\overline 3$ fullerides have been studied for several decades, for example Na$_2$CsC$_{60}$ \cite{Prassides1994} which is
superconducting at ambient pressure with $T_c=12$~K. 
In Na$_2$CsC$_{60}$, 
the distance between the tetrahedral Na$^+$ and nearest C is 2.96 \AA, which is bigger than 
$r_{\mathrm{Na^+}}$+$r_{\mathrm{vdW,C}}=0.95+1.7=2.65$ \AA$\ $ [see Supplementary Material (SM.1)]. 
If however the tetrahedral sites incorporate K, Rb or Cs, or any atoms with a size larger than the size of the hole, these simple geometrical considerations
essentially exclude the existence of such a structure. 

We confirmed the structural instability of Pa$\overline{3}$ Cs$_3$C$_{60}$ at the reported volume by performing a crystal structure relaxation at fixed volume, using the 
Vienna ab initio simulation package (VASP) \cite{vasp_ref1,vasp_ref2,vasp_ref3} (SM. 2). 
Figure~\ref{fig:struct}(i) shows the evolution of the energy difference $\Delta E$ per formula unit between the structure under
relaxation and the final stable FCC structure. We start from the experimentally determined Pa$\overline{3}$ structure \cite{XJChen2208}, 
which turns out to have 
a much higher energy than the FCC structure, with $\Delta E\approx 4.67$ eV. This indicates that the Pa$\overline{3}$ structure is 
energetically unstable. The structural relaxation is accompanied by rotations of the C$_{60}$ molecules and the alignment of the local coordinate systems (see SM.~2). 
The distance between the tetrahedral Cs$^+$ and its nearest C in the relaxed FCC structure is 3.25~\AA, which is still smaller than $r_{\mathrm{Cs}^+} + r_\mathrm{vdW,C} = 3.37$ \AA\, and 
corresponds to a significant pressure of approximately 2.2 GPa at $T=15.4$~K \cite{Ganin2010} (or an even higher pressure at room temperature).  
We note, however, that these results for Cs$_3$Cs$_{60}$ do not imply that all Pa$\overline{3}$ structures are unstable. The Pa$\overline{3}$ Na$_2$CsC$_{60}$ structure reported in Ref.~\onlinecite{Prassides1994} is stable against relaxation, consistent with the above-mentioned fact that Na$^+$ fits well into the tetrahedral hole (see SM.2).  

{\it Large crystal field splitting in the Pa$\overline{3}$ structure.} \
The band structure (black lines) of the unstable Pa$\overline{3}$ structure is shown in 
Fig.~\ref{fig:struct}(c). The tight-binding (TB) Hamiltonian for the bands near the Fermi energy is obtained
from the maximally-localized Wannier functions constructed by wannier90 \cite{wannier90,Pizzi_2020}. These TB bands (orange lines) reproduce the low-energy DFT bands very well. 
We choose orbitals which diagonalize the onsite Hamiltonian 
of each C$_{60}$ molecule. As the four molecules in the unit cell are equivalent to each other in the local 
coordinates, the density of states (DOS) and crystal field  (CF) splittings are the same. As 
shown in Fig.~\ref{fig:struct}(d), the 3-fold degenerate $t_\text{1u}$ orbitals of the isolated molecule 
are split in a ``one down, two up" fashion, with orbital 1 lying below the degenerate orbitals 2 and 3. According to the on-molecule Hamiltonian of the Wannier TB model, the CF splitting in the Pa$\overline{3}$ 
structure is as big as 164 meV. The FCC structure instead has no CF splitting, as shown by the triply degenerate DOS in Fig.~\ref{fig:struct}(h). CF splittings in fullerides tend to suppress superconductivity \cite{Kim2016,Yue2022} since they reduce the local orbital fluctuations  and hence weaken the pairing glue \cite{Yue2021}. 
 
 \begin{table}[b] 
\centering
\caption{Volume dependence of the orbital average of the static on-site
interaction parameters and bandwidth for FCC Cs$_3$C$_{60}$.
The first and second row show the static Coulomb interaction $U_{\mathrm{cRPA}}$ and the 
exchange interaction $J_{\mathrm{cRPA}}$
calculated by cRPA. 
The third and fourth row show the 
phonon-mediated static Coulomb interactions $U_{\mathrm{ph}}$ and on-site exchange interaction $J_{\mathrm{ph}}$ 
extracted from Ref.~\onlinecite{Nomura2015b}. 
The fifth and sixth row show the effective interaction $U(J)_{\mathrm{eff}}=U(J)_{\mathrm{cRPA}}+U(J)_{\mathrm{ph}}$.
The last row shows the bandwidth $W$ of the target bands. The energies are in eV.
}
\label{tab:cRPA_fcc}
\medskip
\begin{tabular}{@{\  }c@{\ \ \ }c@{\ \ \ }c@{\ \ \ }c@{\ \ \ }c@{\  }c@{\  }}
\hline
V (\AA$^3$) &675 & 700  & 724.8  &750 & 775 \tabularnewline
\hline
$U_{\mathrm{cRPA}}$ & 0.6286 & 0.7462& 0.8269 & 0.8872 & 0.9814 \tabularnewline
$J_{\mathrm{cRPA}}$ & 0.0417 & 0.0354& 0.0367 & 0.0372 & 0.0386\tabularnewline
$U_{\mathrm{ph}}$   &-0.17   &-0.16  &-0.1479 &-0.142  &-0.120\tabularnewline
$J_{\mathrm{ph}}$   &-0.05   &-0.05  &-0.05   &-0.05   &-0.051 \tabularnewline
$U_{\mathrm{eff}}$  & 0.5686 & 0.5862& 0.6790 & 0.7452 & 0.8614 \tabularnewline
$J_{\mathrm{eff}}$  &-0.0083 &-0.0146&-0.0133 &-0.0128 &-0.0124 \tabularnewline
$W$		    & 0.570  & 0.528 & 0.474  & 0.429  & 0.367\tabularnewline
\hline
$T_c$ (K)  & $\mathrm{<}7.5$ & $\mathrm{<}7.5$ & 17.7 & 33.7 & 42.5 \tabularnewline
\hline
\end{tabular}
\end{table}

{\it Ab-initio calculations.} \ 
The realistic effective on-molecule Coulomb interactions $U_{\mathrm{eff}}$ and Hund couplings $J_{\mathrm{eff}}$ are shown in the fifth and sixth rows of Tab.~\ref{tab:cRPA_fcc}. They are derived by considering two contributions. The first contribution, $U(J)_{\mathrm{cRPA}}$, is the static interaction obtained by the constrained random phase approximation (cRPA) \cite{cRPA2004}, which considers the screening by electrons in higher-energy bands. The second contribution, $U(J)_{\mathrm{ph}}$, is due to phonon screening and is obtained within  constrained density functional perturbation theory \cite{Nomura2015b}. We calculated $U(J)_{\mathrm{cRPA}}$ for the different volumes using the VASP code \cite{vasp_ref1,vasp_ref2,vasp_ref3} and report the values in the first and second rows of Tab.~\ref{tab:cRPA_fcc}. The parameters $U(J)_{\mathrm{ph}}$ were obtained by linear extrapolation of the data presented in Ref.~\onlinecite{Nomura2015b} and are listed in the third and fourth rows of Tab.~\ref{tab:cRPA_fcc}.
The bandwidth $W$ of the TB bands is shown in the 7th row. 
$J_{\mathrm{eff}}$ is negative for all volumes, which shows that the Hund coupling is inverted due to the overscreening by the phonons. 

We solve the realistic three-band Hubbard model for the $t_{1u}$ orbitals with rotationally invariant Kanamori interactions in the framework of DFT plus dynamical mean-field theory (DMFT) \cite{Georges1996,Kotliar2006}. To deal with the $s$-wave intra-orbital spin-singlet pairing, we implement DFT+DMFT in the Nambu formalism and solve the corresponding three-orbital Anderson impurity model with a 
superconducting bath using continuous-time quantum Monte Carlo simulations based on the hybridization expansion (CT-HYB) \cite{Werner2006,Gull2011}. Four-operator updates are implemented to ensure the ergodicity of the Monte Carlo sampling in the symmetry-broken phase \cite{Semon2014}. Furthermore, the worm algorithm is used to measure the Green's function \cite{Gunacker2015,Yue2022_Worm, Hausoel2023}. The calculations are constrained to paramagnetic and orbitally symmetrized solutions.

\begin{figure}
\includegraphics[clip,width=3.4in,angle=0]{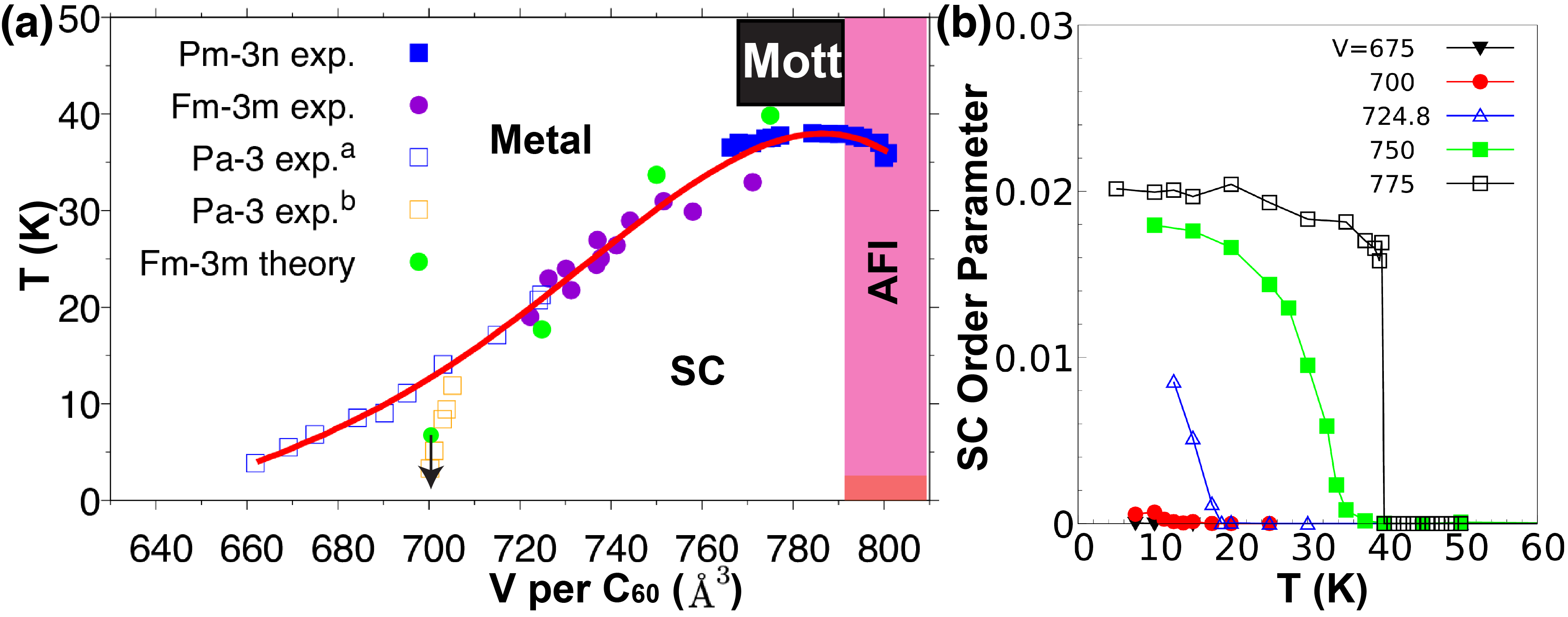}
\caption{
(a) Superconducting phase of Cs$_3$C$_{60}$ as a function of volume per C$_{60}$. The green circles show the $T_c$ of the FCC structure 
obtained using DFT+DMFT in the Nambu formalism. The blue empty squares (with superscript $^a$) and red solid line show the $T_c$ of the Pa$\overline{3}$ structure reported in Ref.~\onlinecite{XJChen2208}. The pink (orange) shading indicates the antiferromagnetic insulator phase for Pm${\overline 3}$n
(Fm${\overline 3}$m) with N\'eel temperature $T_N\sim 46$ (2.2) K.
The light orange empty squares (with superscript $^b$) mark the $T_c$ for Na$_2$Rb$_x$Cs$_{1-x}$C$_{60}$ in the Pa$\overline 3$ structure \cite{Yildirim1995,Brown1999}.
The solid squares and dots show the $T_c$ for the A15 and FCC structure reproduced 
from Refs.~\cite{Ganin2008,Nagamatsu2001,Hebard1991}, respectively.
The black region schematically indicates the Mott insulating solutions at elevated temperature above $T_c$ near $V=775$ \AA$^3$. 
(b) The $T$-dependence of the superconducting order parameter for FCC Cs$_3$C$_{60}$ for the indicated volumes
in the FCC structure.
}
\label{fig:Tc}
\end{figure}

{\it Superconductivity in the stable FCC structure.} \ 
Since the Pa$\overline{3}$ structure is unstable, 
we instead study the superconducting and normal states of FCC Cs$_3$C$_{60}$ under pressure (for compressed volumes). Figure~\ref{fig:Tc}(a) shows the calculated $T_c$ which is determined from the temperature dependence of the superconducting order parameter
as shown in Fig.~\ref{fig:Tc}(b). Our results for $T_c$ well reproduce the experimental $T_c$ for the FCC structure obtained by Ganin {\it et al.} \cite{Ganin2008} in the volume 
range $V \sim 720 - 780$ \AA$^3$. For $V \le 700$ \AA$^3$, no signature of superconductivity is found at the lowest temperature ($\sim 7.5$ K) we can reach in the DFT+DMFT
calculations. The trend of the calculated $T_c$ with decreasing volume is not consistent with the trend reported in Ref.~\onlinecite{XJChen2208} (red solid line) for the Pa$\overline{3}$ structure. On the small volume side, the $T_c<7.5 $~K for $V \sim 700$ \AA$^3$  is instead similar to that of Na$_2$Rb$_x$Cs$_{1-x}$C$_{60}$ in the Pa$\overline{3}$ structure reported in Ref.~\onlinecite{Yildirim1995}. The red solid line at $V < 700$ \AA$^3$ roughly matches the results for K$_3$C$_{60}$ under pressure \cite{Yildirim1995}. 

\begin{figure*}[htp]
\includegraphics[clip,width=0.8\paperwidth,angle=0]{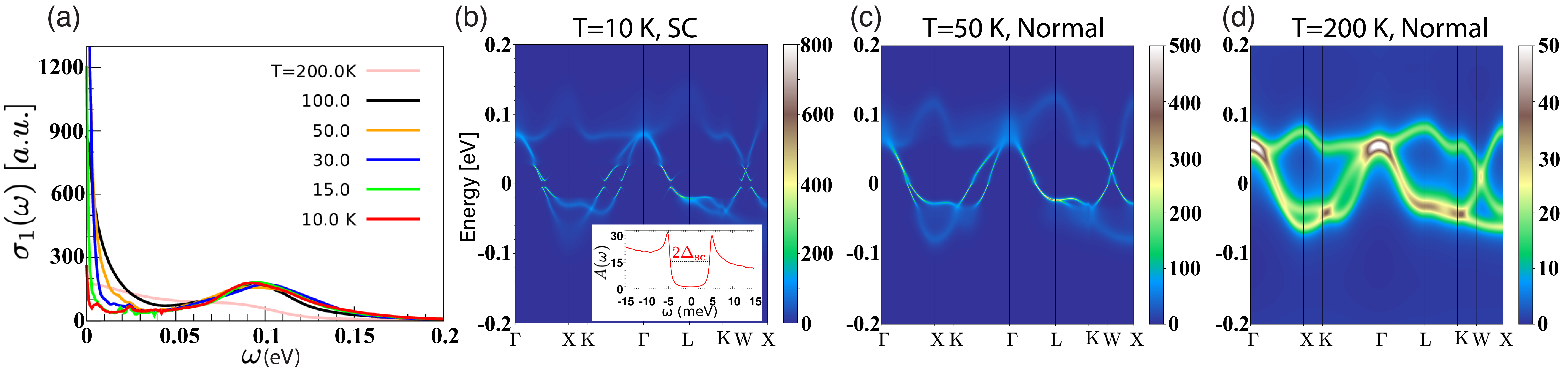}
\caption{Optical conductivity $\sigma_1(\omega)$ (a) and momentum-resolved spectral function $ A({\bf k},\omega)$ (b-d) in a narrow energy window $-0.2 < \omega  < 0.2$ eV at $V=750$ \AA$^3$ at the indicated temperatures. The inset in panel (b) shows the low-energy ${\bf k}$-integrated normal spectrum $A(\omega)$ with $\Delta_{\mathrm{sc}}$ the SC gap.}
\label{fig:optical_akw_V750}
\end{figure*}
\begin{figure*}[htp]
\includegraphics[clip,width=0.8\paperwidth,angle=0]{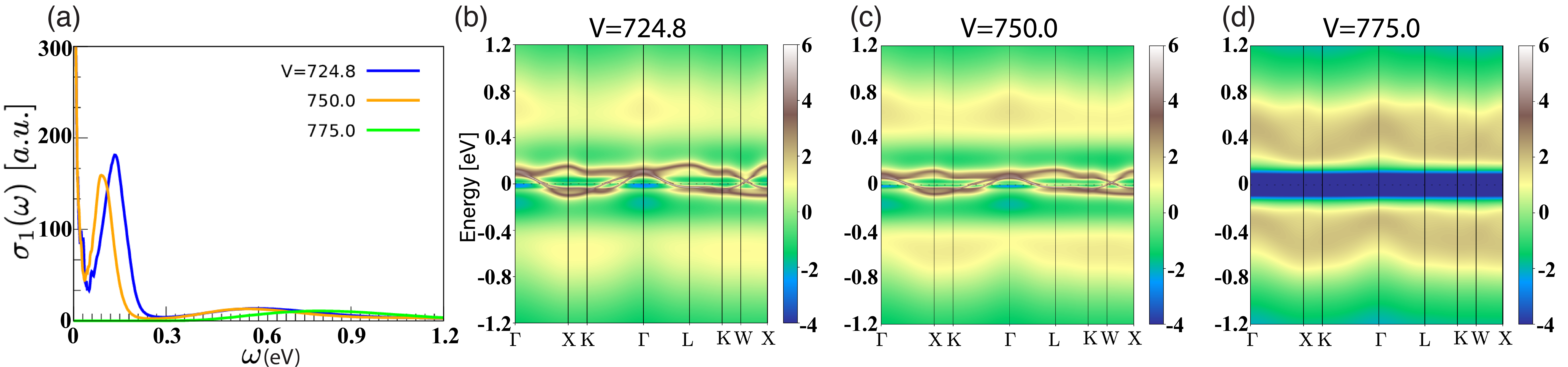}
\caption{Optical conductivity (a) and momentum-resolved spectral function $ A({\bf k},\omega)$ (b-d) in a wide energy window $-1.2 < \omega < 1.2 $ eV, for the normal state at $T=50$ ($>T_c$). 
The volumes are indicated in each panel. In (b-d), we show the intensity on a log scale 
for a better visualization of the Hubbard bands.}
\label{fig:optical_akw_T50K}
\end{figure*} 

A noteworthy observation is that the transition from the $s$-wave unconventional superconductor to the metallic state is second order, while the transition to the Mott insulator is discontinuous.  
This can be seen from $\Delta(T)$, the temperature dependence of the superconducting order parameter. At $V=750$ \AA$^3$, where the system is moderately correlated,
$\Delta(T)$ follows the relation
$\Delta(T)\propto|T-T_c|^{1/2}$ near the critical point, and we use this relation to determine $T_c$. At  $V=775$ \AA$^3$, in the strong correlation regime,  $\Delta(T)$ drops abruptly to 0 as the system enters the Mott insulating phase. 
The first order transition from superconductor to Mott insulator seen in Cs$_3$C$_{60}$ on the strong coupling side of the dome is 
similar to that between the $d$-wave superconductor and Mott insulator reported in Refs.~\onlinecite{Lefebvre2000} and \onlinecite{Sordi2012}. 

{\it Momentum-resolved spectra and optical conductivity.} \ 
The momentum-resolved spectrum can be calculated from the Green's function $G$ as $A({\bf k},\omega)=-2 \sum_{\alpha=1}^{3} \mathrm{\ Im} G_{\alpha\uparrow,\alpha\uparrow}({\bf k},\omega)/\pi$, 
where the factor of 2 comes from the spin degeneracy.
In the superconducting state, to get the real frequency lattice Green's function $G({\bf k},\omega)$, one has to perform an analytic continuation from the Matsubara to the real frequency axis not only of the normal self-energy $\Sigma^{\text{nor}}$, but also of the anomalous self-energy $\Sigma^{\text{ano}}$.  
Here, we adopt the recently developed maximum entropy analytic continuation method for $\Sigma^{\text{ano}}$ \cite{Yue2023}. In this approach one defines an auxiliary self-energy with positive-definite spectral weight to compute the real-frequency $\Sigma^{\text{ano}}$. 
We calculate the optical conductivity $\sigma_1(\omega)$ within the framework of linear response and in the long-wave-length limit \cite{Tomczak2009,Wissgott2012}. In the superconducting state, one has to express the dipole matrix elements and matrix spectral function \cite{Assmann2016} in the Nambu formalism. 

We first consider the temperature dependence of $A({\bf k},\omega)$ and $\sigma_1(\omega)$ at fixed volume. Here, we choose the volume $V=750$ \AA$^3$ corresponding 
to moderate correlations ($U_{\mathrm{eff}}/W\approx 1.74$) and $T_c=33.7$~K located slightly to the left of the maximum of the SC dome. The results are shown in Fig.~\ref{fig:optical_akw_V750}. 
 At high temperature, $T=200$ K, the system is not very coherent, as shown in panel (d). 
 The bandwidth is renormalized with respect to the noninteracting $W$ by about a factor of three, consistent with the mass enhancement $m^*/m\approx 1-\mathrm{Im}\Sigma^{\mathrm{nor}}(i\omega_{0})/\omega_{0}=3.3$, where $\omega_0=\pi T$ is the first Matsubara frequency.
 Close to, but above $T_c$, the quasi-particle bands become sharper, as shown in panel (c). When the temperature drops below $T_c$, a superconducting gap opens at all momenta. At $T=10$ K ($< T_c/3$, panel (b)), one observes a gap of 4.6 meV (estimated from the ${\bf k}$-integrated normal spectra, see inset) and back-bending of the Bogolyubov quasi-particle bands. 
 
The temperature dependence of the real part of the conductivity, $\sigma_1(\omega)$, is shown in panel (a). 
 As one decreases $T$ from 100 K (black line in panel (a)) to 50 K (orange line), the Drude peak of $\sigma_1(\omega)$ grows in height but narrows, which results in a loss of spectral weight in the energy range from 5 meV to 50 meV. 
 At $T=30$~K (blue line), the system enters the
 SC phase, marked by additional spectral weight loss below 30 meV, while a narrow Drude feature still persists. As one decreases $T$  further from 
 $T=15$~K (green) to 10 K (red), the Drude peak becomes very small and sharp. 
 The persistent Drude peak at $T=10$ K in a very low-energy range ($\omega < 5 $ meV) 
 is related to the faint spectral weight in $A(\bf{k},\omega)$ at the Fermi energy, which we have also previously observed in simulations of K$_3$C$_{60}$ \cite{Yue2023}. This faint spectral weight indicates the break-up of some Cooper pairs at nonzero temperature. The calculated $\sigma_1(\omega)$ is
 qualitatively consistent with the conductivity measurements for K$_3$C$_{60}$ and  Rb$_3$C$_{60}$ \cite{Degiorgi1994}. 

Figure~\ref{fig:optical_akw_T50K} shows the normal state ($T=50$~K $>T_c$) $A({\bf k},\omega)$ (panels (b-d)) and $\sigma_1(\omega)$ (panel (a)) at different volumes. The first observation
is that there is no Drude peak  in $\sigma_1(\omega)$ for $V=775$ \AA$^3$ (green line), since at this volume the system is a Mott insulator with a fully gapped spectral function (panel (d)) above $T_c$. (This system is in the strong coupling regime with $U_{\mathrm{eff}}/W\approx 2.35$.) 
The broad hump located around $\omega\sim 0.8 $ eV arises from transitions between the Hubbard bands located at $\omega\sim \pm U_{\mathrm{eff}}/2$ 
(panel (d), $U_{\mathrm{eff}}\approx 0.86$ eV) and from transitions between the quasi-particle bands and the Hubbard bands \cite{footnote_sigma}. As one decreases the volume by increasing pressure, a Mott insulator 
to metal transition is triggered and the system becomes more metallic, as seen from panels (c) ($V=750$ \AA$^3$) and (b) ($V=724.8$ \AA$^3$). The weight of the Hubbard bands is reduced and the width of the quasi-particle bands 
increases, as can be seen by 
by comparing panels (b) and (c). As a result
of this increasing bandwidth, the peak of $\sigma_1(\omega)$ in the energy range $0.06<\omega<0.3$ eV broadens and shifts to larger energies. 

{\it Discussion and Conclusions.} \ 
We pointed out that the recently proposed Pa$\overline{3}$ structure of Cs$_3$C$_{60}$ has an unrealistic spacing between the C$_{60}$ molecules and Cs dopants and is structurally very unstable according to DFT calculations. Even if the Pa$\overline{3}$ structure could be somehow realized at ambient pressure, this system would feature a large crystal field splitting between the $t_\text{1u}$ bands, and thus most likely not exhibit superconductivity. 

Since the Pa$\overline{3}$ structure relaxes to the stable FCC structure in DFT, 
we studied the electronic properties of the FCC structure using state-of-the-art DFT+DMFT simulations in the Nambu-formalism. 
The superconducting order parameter, critical temperature $T_c$, as well as the momentum-resolved spectra and optical conductivity in the SC
phase and normal phase were calculated. Our calculated $T_c$ values as a function of volume are consistent with previous experiments on the FCC structure. The transition between the $s$-wave SC phase and the Mott insulator at large volumes is found to be first order. The optical conductivity features a small Drude peak at the lowest temperature $T=10$ K that we can reach, 
which is also consistent with optical measurements. This peak can be related to the faint spectral weight at the Fermi energy and the coexistence of paired and unpaired electrons in this small-gap superconductor. 

Our study does not rule out the possibility of superconductivity in other fullerides with Pa$\overline{3}$ structure, such as Na$_2$Rb$_x$Cs$_{1-x}$C$_{60}$ \cite{Prassides1994,Yildirim1995}. We confirmed that Na$_2$CsC$_{60}$ is stable according to the DFT relaxation. It would be interesting to study the effect of the orientation of the C$_{60}$ molecules on $T_c$. Since the symmetry is
lowered compared to the FCC structure, the resulting crystal field splitting is expected to suppressed local orbital fluctuations and lower $T_c$,
compared to the FCC structure with the same volume. 

{\it Acknowledgements. ---}\
The DMFT calculations were performed on the Beo05 and Beo06 cluster at the University of Fribourg, using a code based on iQIST \cite{HUANG2015140,iqist}. C.Y. and P.W. acknowledge support from SNSF Grant No. 200021-196966. 
Y.N. acknowledge support from Grant-in-Aids for Scientific Research (JSPS KAKENHI) [Grant Nos. JP23H04869, JP23H04519, JP23K03307, and JP21H01041], MEXT as ``Program for Promoting Researches on the Supercomputer Fugaku'' (Grant No. JPMXP1020230411), and JST (Grant No. JPMJPF2221). KP acknowledges financial support by Grants-in-Aid for Scientific Research (JSPS KAKENHI Grant Numbers JP21H01907 and JP22K18693).


\begin{thebibliography}{99}
\bibitem{Gunnarsson1997} O. Gunnarsson, Rev. Mod. Phys. \textbf{69}, 575 (1997).
\bibitem{Capone2009} Massimo Capone, Michele Fabrizio, Claudio Castellani, and Erio Tosatti, Rev. Mod. Phys. \textbf{81}, 943 (2009).
\bibitem{PalstraCs3C60Tc40K} Palstra, T. T. M., O. Zhou, Y. Iwasa, P. E. Sulewski, R. M.  Fleming, and B. R. Zegarski. Solid State Commun. \textbf{93}, 327 (1995). 
\bibitem{Ganin2008} A. Y. Ganin, Y. Takabayashi, Y. Z. Khimyak, S. Margadonna, A. Tamai, M. J. Rosseinsky, K. Prassides.  Nat. Mater. \textbf{7}, 367 (2008).
\bibitem{Takabayashi2009} Y. Takabayashi, A. Y. Ganin, P. Jeglič, D. Arčon, T. Takano, Y. Iwasa, Y. Ohishi, M. Takata, N. Takeshita, K. Prassides, M. J. Rosseinsky. Science \textbf{323}, 1585 (2009)
\bibitem{Zadik2015} R. H. Zadik, Y. Takabayashi, G. Klupp, \textit{ et al.}, Sci. Adv. \textbf{1}, e1500059 (2015).
\bibitem{Crespi2002} J. E. Han, O. Gunnarsson, and V. H. Crespi, Phys. Rev. Lett. \textbf{90}, 167006 (2003).
\bibitem{Capone2002} M. Capone, M. Fabrizio, C. Castellani, and E. Tosatti, Science \textbf{296}, 2364 (2002).
\bibitem{Nomura2012} Y. Nomura, K. Nakamura, and R. Arita, Phys. Rev. B \textbf{85}, 155452 (2012).

\bibitem{Nomura2015} Y. Nomura, S. Sakai, M. Capone, and R. Arita, Sci. Adv. \textbf{1}, e1500568 (2015).
\bibitem{Hoshino2017} S. Hoshino, and P. Werner, Phys. Rev. Lett. \textbf{118}, 177002 (2017).
\bibitem{Yue2020} C. Yue, S. Hoshino, and P. Werner, Phys. Rev. B \textbf{102}, 195103 (2020).
\bibitem{Hoshino2015} S. Hoshino, and P. Werner,  Phys. Rev. Lett. \textbf{115}, 247001 (2015).
\bibitem{Werner2016} P. Werner, S. Hoshino, and H. Shinaoka,  Phys. Rev. B \textbf{94}, 245134 (2016).
\bibitem{XJChen2208}  D. Peng, R.-S. Wang, L.-N. Zong, X.-J. Chen. arXiv \textbf{2208}, 09429 (2022).
\bibitem{Zhou1992} O. Zhou, G. B. M. Vaughan, Q. Zhu, J.E. Fischer, P. A. Heinney, N. Coustel, John P. McCauley, Jr., and A. B. Smith III, Science \textbf{255}, 833 (1992).
\bibitem{Prassides1994} K. Prassides, C. Christides, I. M. Thomas, J. Mizuki, K. Tanigaki, I. Hirosawa, T. W. Ebbesen Science \textbf{263}, 960 (1994).
\bibitem{Heiney1991_PRL} P. A. Heiney, J. E. Fisher, A. R. McGhie, W. J. Romanow, A. M. Denenstein, J. P. McCauley Jr., A. B. Smith, and D. E. Cox. Phys. Rev. Lett. \textbf{66}, 2911 (1991)
\bibitem{footnote_disorder} Experimentally, the known fulleride compounds show orientational or merohedral disorder. Such type of disorder is not considered in our calculations. 
\bibitem{vasp_ref1} G. Kresse, and J. Hafner, {Phys. Rev. B} \textbf{47}, 558 (1993).
\bibitem{vasp_ref2} G. Kresse, and J. Furthm\"uller, {Comput. Mater. Sci.} \textbf{6}, 15 (1996).
\bibitem{Ganin2010} A. Y. Ganin, Y. Takabayashi, P. Jeglic, D. Arcon, A. Potocnik, P. J. Baker, Y. Ohishi, M. T. McDonald, M. D. Tzirakis, A. McLennan, G. R. Darling, M. Takata, M. J. Rosseinsky, and K. Prassides, Nature \textbf{466}, 221 (2010). 
\bibitem{vasp_ref3} G. Kresse, and J. Furthm\"uller, {Phys. Rev. B} \textbf{54}, 11169 (1996).
\bibitem{wannier90} A. A. Mostofi, J. R. Yates, G. Pizzi, Y.-S. Lee, I. Souz, D. Vanderbilt, and N. Marzari, {Comput. Phys. Commun.} \textbf{185}, 2309 (2014).
\bibitem{Pizzi_2020} G. Pizzi, V. Vitale, R. Arita \textit{ et al.}, {J. Phys.: Condens. Matter} \textbf{32}, 165902 (2020).
\bibitem{Kim2016} M. Kim, Y. Nomura, M. Ferrero, P. Seth, O. Parcollet, and A. Georges, Phys. Rev. B {\bf 94}, 155152 (2016).
\bibitem{Yue2022} C. Yue, Y. Nomura, and P. Werner, Phys. Rev. Lett. \textbf{129}, 066403 (2022).
\bibitem{Yue2021} C. Yue, S. Hoshino, A. Koga, and P. Werner, Phys. Rev. B \textbf{104}, 075107 (2021).
\bibitem{cRPA2004} F. Aryasetiawan, M. Imada, A. Georges, G. Kotliar, S. Biermann, and A. I. Lichtenstein. Phys. Rev. B \textbf{70}, 195104 (2004)
\bibitem{Nomura2015b} Y. Nomura, and R. Arita, {Phys. Rev. B} \textbf{92}, 245108 (2015).
\bibitem{Georges1996} A. Georges, G. Kotliar, W. Krauth, and M. J. Rozenberg, {Rev. Mod. Phys.} \textbf{68}, 13 (1996).
\bibitem{Kotliar2006} G. Kotliar, S. Y. Savrasov, K. Haule, V. S. Oudovenko, O. Parcollet, and C. A. Marianetti, Rev. Mod. Phys. \textbf{78}, 86 (2006).
\bibitem{Werner2006} P. Werner, A. Comanac, L. de’ Medici, M. Troyer, and A. J. Millis, {Phys. Rev. Lett.} \textbf{97}, 076405 (2006).
\bibitem{Gull2011} E. Gull, A. J. Millis, A. I. Lichtenstein, A. N. Rubtsov, M. Troyer, and P. Werner, {Rev. Mod. Phys.} \textbf{83}, 349 (2011).
\bibitem{Semon2014} P. S\'emon, G. Sordi, and A.-M. S. Tremblay, Phys. Rev. B \textbf{89}, 165113 (2014).
\bibitem{Gunacker2015} P. Gunacker, M. Wallerberger, E. Gull, A. Hausoel, G. Sangiovanni, and K. Held, Phys. Rev. B {\bf 92}, 155102 (2015).
\bibitem{Yue2022_Worm} C. Yue, H. Aoki, and P. Werner, Phys. Rev. B \textbf{106}, L180506 (2022).
\bibitem{Hausoel2023} A. Hausoel, M. Wallerberger, J. Kaufmann, K. Held and G. Sangiovanni. arXiv:2211.06266 (2022).
\bibitem{Yildirim1995} T. Yildirim, J. E. Fischer, R. Dinnebier,  P. W. Stephens, and C. L. Lin, Solid State Commun. \textbf{93}, 269 (1995).
\bibitem{Brown1999} C. M. Brown, T. Takenobu, K. Kordatos, K. Prassides, Y. Iwasa, K. Tanigaki, Phys. Rev. B \textbf{59}, 4439 (1999).
\bibitem{Nagamatsu2001} Nagamatsu, J., Nakagawa, N., Muranaka, T., Zenitani, Y. Akimitsu, J. Superconductivity at 39 K in magnesium diboride. Nature \textbf{410}, 63 (2001).
\bibitem{Hebard1991} Hebard, A. F., Rosseinsky, M. J., Haddon, R. C., Murphy, D. W., Glarum, S. H., Palstra, T. T. M., Ramirez, A. P. and Kortan, A. R. Nature \textbf{350}, 600 (1991).
\bibitem{Lefebvre2000} S. Lefebvre, P. Wzietek, S. Brown, C. Bourbonnais, D. Jerome, C. Meziere, M. Fourmigue, and P. Batail, Phys. Rev. Lett. \textbf{85}, 5420 (2000).
\bibitem{Sordi2012} G. Sordi, P. S\'emon, K. Haule, and A.-M. S. Tremblay, Phys. Rev. Lett. \textbf{108}, 216401 (2012).
\bibitem{Yue2023} C. Yue and P. Werner, arXiv:2203, 16888 (2023).
\bibitem{Tomczak2009} J.M. Tomczak and S. Biermann, Phys. Rev. B \textbf{80}, 085117 (2009).
\bibitem{Wissgott2012} P. Wissgott, J. Kuneš, A. Toschi and K. Held, Phys. Rev. B \textbf{85}, 205133 (2012).
\bibitem{Assmann2016} E. Assmann, P. Wissgott, J. Kune{\v s}, A. Toschi, P. Blaha, K. Held, Comput.  Phys. Commun. \textbf{202}, 1 (2016). 
\bibitem{Degiorgi1994} L. Degiorgi, G. Briceno, M. S. Fuhrer, A. Zettl, and P. Wachter, Nature \textbf{369}, 541-543 (1994).
\bibitem{footnote_sigma} Note that only the $t_{1u}$ orbitals are included in our DMFT simulations. In experiments, excitations to other bands could also occur at these energies. 
\bibitem{HUANG2015140} L. Huang, Y. Wang, Z. Y. Meng, L. Du, P. Werner, and X. Dai, {Comput. Phys. Commun} \textbf{195}, 140 (2015).
\bibitem{iqist} L. Huang, {Comput. Phys. Commun} \textbf{221}, 423 (2017).
\end{thebibliography}
\end{document}


\title{\textbf{Supplementary Information} \\ Instability of Pa$\overline 3$ Cs$_{3}$C$_{60}$ at ambient pressure and superconducting state \\ of the FCC phase}

\author{Changming Yue}
\email{changming.yue@unifr.ch}
\affiliation{Department of Physics, University of Fribourg, 1700 Fribourg, Switzerland}

\author{Yusuke Nomura}
\affiliation{Department of Applied Physics and Physico-Informatics, Keio University, 3-14-1 Hiyoshi, Kohoku-ku, Yokohama, 223-8522, Japan}

\author{Kosmas Prassides}
\affiliation{Department of Physics, Graduate School of Science, Osaka Metropolitan University, Osaka 599-8531, Japan}

\author{Philipp Werner}
\email{philipp.werner@unifr.ch}
\affiliation{Department of Physics, University of Fribourg, 1700 Fribourg, Switzerland}

\maketitle

\beginsupplement

\section*{SM1. Smallest distance between tetrahedral C$\text{s}$ and C  in P$\text{a}$$\overline{3}$ C$\text{s}$$_{3}$C$_{60}$}
\begin{figure*}[htp]
\includegraphics[clip,width=0.4\paperwidth,angle=0]{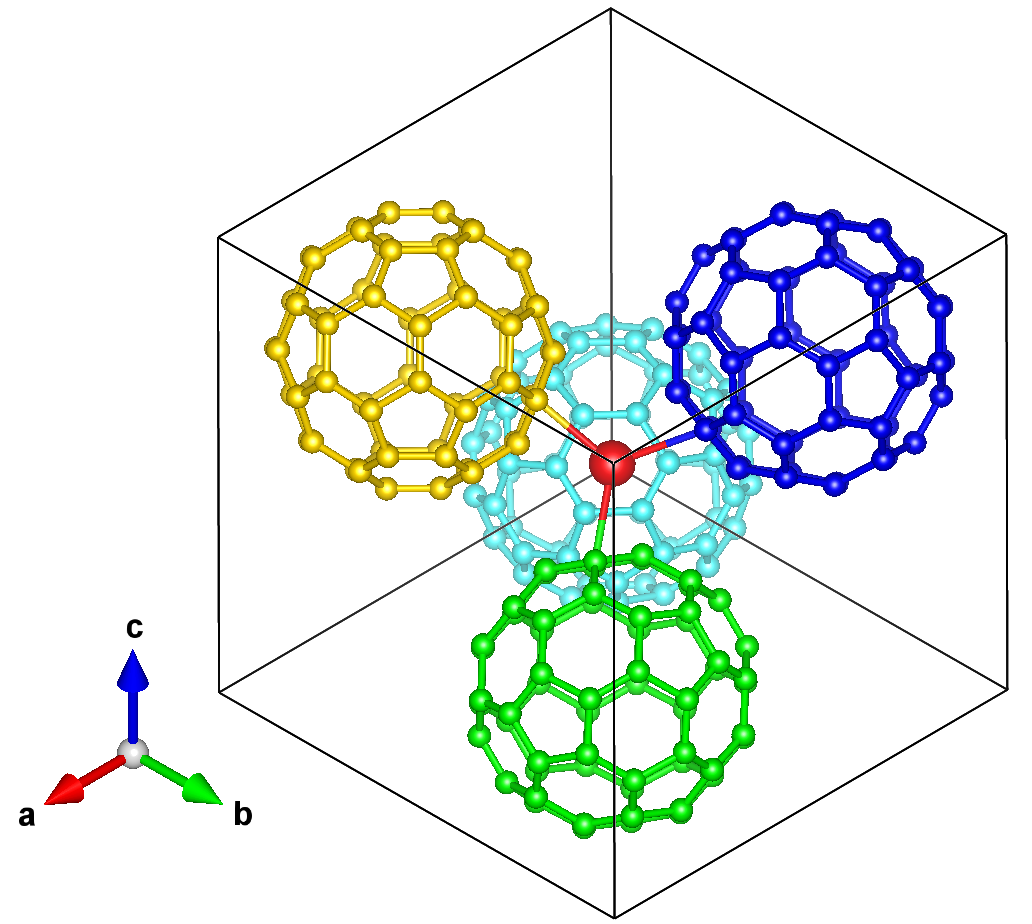}
\caption{The coordination environment of tetrahedral Cs in Pa$\overline 3$ Cs$_{3}$C$_{60}$ viewed along the $C_3$ symmetric axis [$\overline{1}\overline{1}\overline{1}$].
For a better visualization, we use different colors to distinguish the C$_{60}$ molecules with different orientations (as in the main text).}
\label{fig:tetrahedraCs}
\end{figure*}

Figure~\ref{fig:tetrahedraCs} shows the coordination environment of tetrahedral Cs (red dot) in Pa$\overline 3$ Cs$_{3}$C$_{60}$ reported in Ref.~\onlinecite{XJChen2208}.
The smallest distances between the nearest carbon atom and the tetrahedral Cs are shown by three equivalent ``bonds'' connecting this Cs and three nearby C$_{60}$ molecules colored in 
gold, green and blue, respectively. 
The  distance is 2.70 \AA, which is significantly smaller than the sum of the ionic radius of Cs$^+$ ($r_{\mathrm{Cs^+}}=1.67$~\AA) and the van der Waals radius of C ($r_{\mathrm{vdW,C}}=1.7$ \AA). For this reason, it is unlikely that this structure is stable at ambient pressure. 

\section*{SM2. Structural Relaxation}

The density functional theory (DFT) calculations, including the structural relaxation and band structure calculations, are performed using 
the Vienna ab initio simulation package (VASP) code with the Perdew-Burke-Ernzerhof exchange-correlation functional
 \cite{vasp_ref1,vasp_ref2,vasp_ref3} and the cutoff energy 500 eV. We sample the Brillouin zone using
$7\times7\times7$ mesh grids for the DFT calculations and $5\times5\times5$ mesh grids for the constrained random phase approximation calculation.

\begin{figure*}[t]
\includegraphics[clip,width=0.8\paperwidth,angle=0]{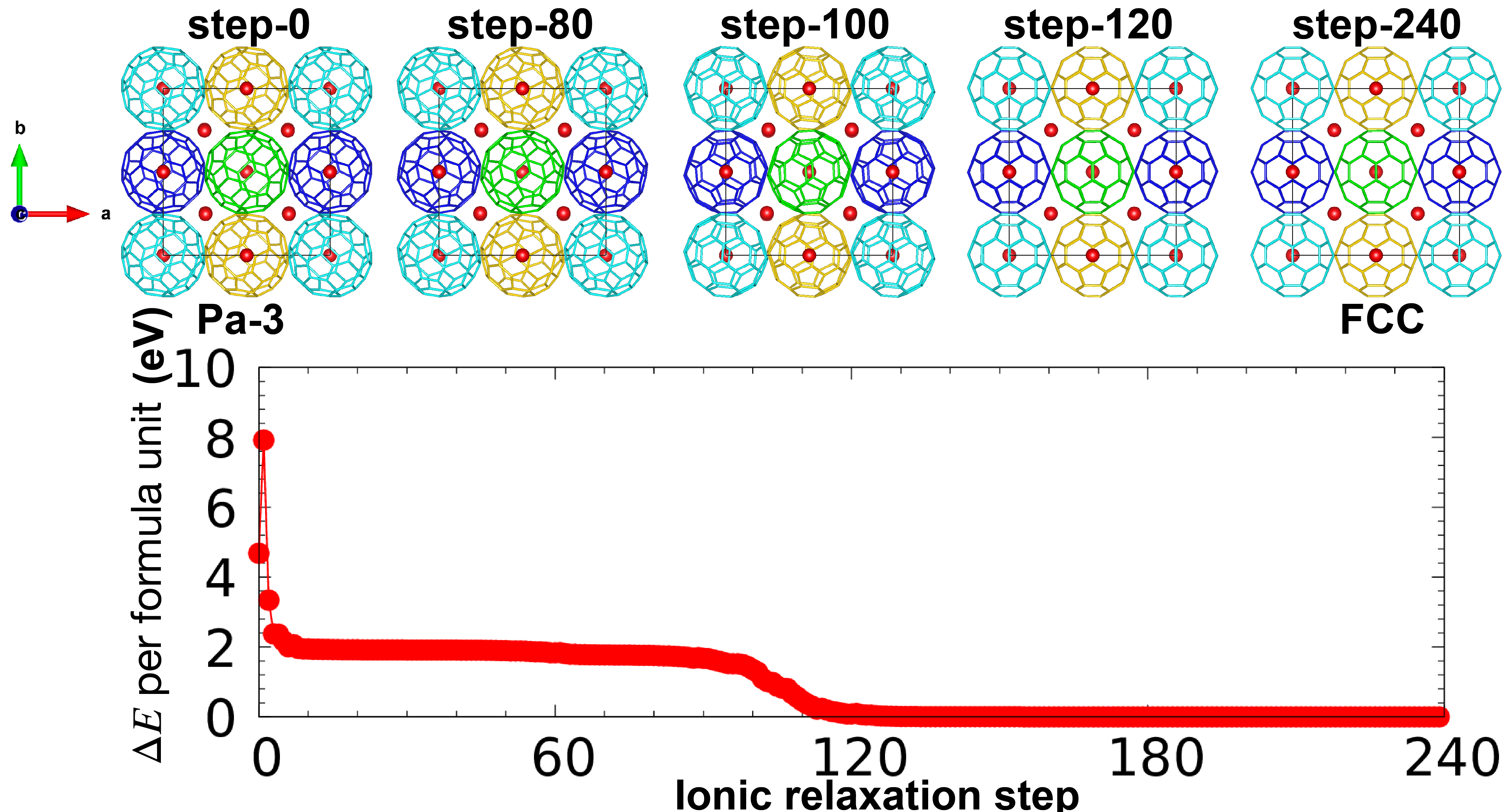}
\caption{Top panel: Evolution of the structure of Cs$_{3}$C$_{60}$ during the relaxation starting from the space group Pa$\overline{3}$.
 Lower panel  (same as Fig.~2(i) in the main text):
Energy difference $\Delta E$ per formula unit between the structure under relaxation and the final stable FCC structure as a function of ionic relaxation steps. 
}
\label{fig:Cs3C60_relax}
\end{figure*}

The structural relaxation is performed with fixed volume.
Figure~\ref{fig:Cs3C60_relax} shows how the structure of Cs$_{3}$C$_{60}$ evolves 
during the relaxation starting from the Pa$\overline 3$ structure reported in Ref.~\onlinecite{XJChen2208}. As we can see from the top
panel, the C$_{60}$ molecules are rotated during the relaxation, especially between step 80 and step 120. In the relaxed structure, there is no difference 
in the orientations of the C$_{60}$ molecules and the system corresponds to the space group Fm$\overline{3}$ (FCC). 

\section*{SM3. N$\text{a}$$_2$C$\text{s}$C$_{60}$ }

\begin{figure*}[htp]
\includegraphics[clip,width=0.8\paperwidth,angle=0]{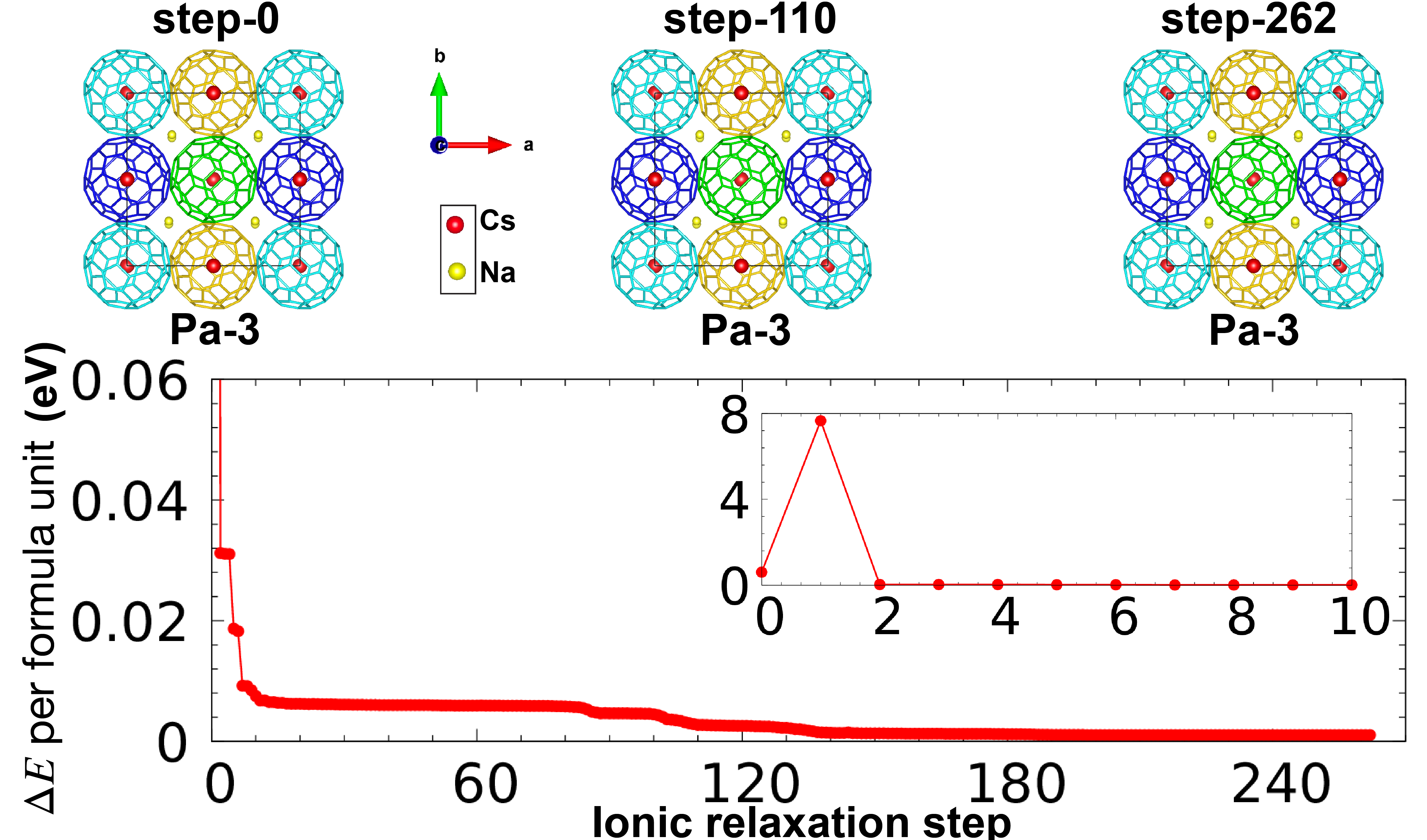}
\caption{Top panel: Evolution of the structure of Na$_{2}$CsC$_{60}$ during relaxation starting from the space group Pa$\overline{3}$.
 Lower panel: 
Energy difference $\Delta E$ per formula unit between the structure under relaxation and the final stable structure as a function of ionic relaxation steps. The inset shows 
$\Delta E$ for the first 11 steps on a larger $y$-range. 
}
\label{fig:Na2CsC60_relax}
\end{figure*}

Na$_2$CsC$_{60}$ crystallizes in the space group Pa$\overline{3}$. We relax this system starting from the main orientation of the structure reported in Ref.~\cite{Prassides1994}. 
As shown by the lower panel of Fig.~\ref{fig:Na2CsC60_relax}, the energy gain after relaxation is 0.6 eV per formula unit, which is much smaller than in the case of Pa$\overline{3}$ Cs$_{3}$C$_{60}$ after relaxation (4.67 eV). Furthermore, the structure is only changed slightly during the relaxation 
and the space group remains Pa$\overline{3}$. Hence, Pa$\overline{3}$ Na$_2$CsC$_{60}$ reported in Ref.~\cite{Prassides1994} is a stable structure, in contrast to Pa$\overline{3}$ Cs$_3$C$_{60}$ reported in Ref.~\cite{XJChen2208}.

